\documentclass[aps,prd,superscriptaddress,showpacs,nofootinbib,notitlepage,twocolumn]{revtex4-1}
\usepackage{graphicx,subfigure,amsmath,amsfonts,amssymb,multirow,mathrsfs}
\usepackage{ulem,soul}
\usepackage[colorlinks,linkcolor=blue,citecolor=blue,urlcolor=blue]{hyperref}
\usepackage[dvipsnames,svgnames]{xcolor}

\newcommand{\DM}{\mathrm{DM}}
\newcommand{\dd}{\mathrm{d}}

\graphicspath{{./}{figures/}}

\begin{document}

\title{First $\Omega_{\rm m}-w$ Measurement Using Only Fast Radio Burst Observations}
\author{Yi-Ying Wang}
\affiliation{Purple Mountain Observatory, Chinese Academy of Sciences, Nanjing 210023, People's Republic of China}
\author{Shi-Jie Gao}
\affiliation{School of Astronomy and Space Science, Nanjing University, Nanjing, 210023, People's Republic of China}
\affiliation{Key Laboratory of Modern Astronomy and Astrophysics, Nanjing University, Ministry of Education, Nanjing, 210023, People's Republic of China}
\author{Chi Zhang}
\affiliation{Purple Mountain Observatory, Chinese Academy of Sciences, Nanjing 210023, People's Republic of China}
\author{Yi-Zhong Fan}
\email{The corresponding author: yzfan@pmo.ac.cn}
\affiliation{Purple Mountain Observatory, Chinese Academy of Sciences, Nanjing 210023, People's Republic of China}
\affiliation{School of Astronomy and Space Science, University of Science and Technology of China, Hefei, Anhui 230026, People's Republic of China}

\begin{abstract}
Though fast radio bursts (FRBs) are widely advocated as a dark energy probe, no valid $\Omega_m-w$ constraint has been obtained exclusively from FRB data to date. Benefiting from a rapidly growing localized sample and a novel statistical framework for modeling the dispersion measure (DM), we report the first simultaneous inference of the matter density $\Omega_m$ and dark energy equation of state parameter $w$ from FRBs alone, without external observational priors on these two parameters.
We combine a multi-parameter emulator for the nonlinear matter power spectrum with a physically motivated electron prescription to construct a cosmology-dependent  cosmic-DM distribution.  
Within this unified analysis framework, the mean and variance of DM respond self-consistently to the same underlying cosmology, directly connecting the FRB likelihood to both cosmic expansion and structure growth. Analyzing 125 localized FRBs at $0.105\leq z\leq2.148$, we obtain $\Omega_m = 0.33^{+0.10}_{-0.11}$ and $w = -0.80^{+0.28}_{-0.38}$ under the flat $w$CDM model. We find a new positive $\Omega_m-w$ degeneracy in the FRB posterior, opposite to the strong anticorrelation derived from Type Ia supernovae, offering a complementary test of dark energy properties and highlighting their potential in future precision cosmological studies. 
\end{abstract}
\maketitle

\section{Introduction}\label{sec:introduction}

Dark energy (DE) drives the accelerated expansion of the Universe, but its physical nature remains unknown. Constraints on its equation of state (EoS) from individual cosmological probes are limited by parameter degeneracies, motivating joint analyses of complementary observables \citep{Weinberg2013}. Recent DESI DR2 baryon acoustic oscillations (BAO) measurements, combined with cosmic microwave background (CMB) observations and Type Ia supernovae (SNe Ia), favor evolving DE over a cosmological constant, with a significance that depends on the adopted SN sample \citep{DESI2025}. Independent probes with different parameter degeneracies are therefore essential for testing this interpretation. Fast radio bursts (FRBs), a population of brief, intense radio outbursts of cosmological origin \citep{Lorimer2007Sci,Petroff2022A&AR,Lorimer2024}, can serve as such a probe through the relation between their dispersion measures (DMs) and host-galaxy redshifts, as first proposed by \citet{Zhou2014}. 
DM traces the line-of-sight free-electron column and thus provides a new way to measure the path length traveled by radio photons.
One advantage is that FRBs have parameter sensitivities complementary to those of BAO, providing a means of breaking degeneracies in the $\Omega_m-w$ plane \cite{Zhou2014}. 

Forecasts have highlighted the promise of using FRBs as a cosmological probe~\cite{Zhou2014,Gao2014,Walters2018ApJ,Wei2018,2019PhRvD.100h3533K,Zhang:2023gye,SKATransientsSWG:2026isd}, but jointly constraining $\Omega_m$ and $w$ with FRBs alone remains challenging, because the localized sample is still limited and the dispersion-measure (DM) contributions from source environments, host galaxies, and intervening structure remain uncertain. Existing analyses have therefore generally adopted external priors on the matter density or combined FRBs with other cosmological probes to constrain $w$~\cite{Yan2025ChPhC,Ribeiro2026arXiv}. A further difficulty arises from the modeling of the DM distributions themselves. Many analyses adopt simulation-calibrated cosmic or host probability density functions (PDFs)~\cite{WQ2022,LY2023ApJ,Wang2025,GDH2025A&A,ZhangJG2026arXiv},
including the cosmic-DM distribution $p_{\rm cos}$ calibrated with TNG300~\cite{Zhang2021cosmic} and host-DM distributions derived from TNG100~\cite{Zhang2020host} \footnote{Both runs belong to IllustrisTNG~\cite{Weinberger2017}, a suite of cosmological gravito-magnetohydrodynamical simulations of galaxy formation that adopts a fiducial $\Lambda$CDM cosmology and includes radiative cooling and feedback from stars and active galactic nuclei.}. 
These calibrations provide physically motivated descriptions of DM fluctuations, but a cosmic-DM PDF calibrated at a single fiducial cosmology does not capture how electron clustering responds to changes in $\Omega_m$ or in the DE sector. Cosmic-DM scatter depends on both cosmology and baryonic feedback, and standard fluctuation-amplitude parameterizations do not reproduce the full redshift and feedback dependence found in hydrodynamic simulations~\cite{Sharma2025ApJ}. Exploiting the cosmological information in FRB data therefore requires a likelihood that consistently propagates cosmology into both the mean DM--redshift relation~\cite{Macquart2020} and its scatter, while accounting for the uncertain host contribution.

To circumvent these existing obstacles, we relate the electron power spectrum $P_{ee}$ to the variance of cosmic DM $\sigma^2_{\rm DM_{\rm cos}}$ via the established formalism of Ref.~\cite{McQuinn2014ApJ}. Using the \texttt{GokuNEmu} emulator~\cite{GokuEmu2025,Yang2026}, which fully covers the dynamical DE parameter space favored by the latest DESI constraints, we self-consistently derive the mean and variance of $\DM_{\rm cos}$ for the same cosmology, while jointly inferring the host-FRB population parameters. Applying this new framework to localized FRBs, we report the first valid $\Omega_m-w$ contour derived exclusively from FRB data and a positive $\Omega_m$--$w$ degeneracy, without informative external observational priors on $\Omega_m$ and $w$.

\section{Dispersion measure modeling}\label{sec:methods}
For an FRB at source redshift $z_s$, the observed DM is given by the integral of the electron number density $n_e$ along the line of sight, 
\begin{equation}
\DM_{\rm obs}({\bf r},z) = \int^{z_s}_0\frac{n_e({\bf r},z)c{\rm d}z}{(1+z)^2H(z)},
\end{equation}
which can be decomposed as 
\begin{equation}\label{eq:dm-decomposition}
\begin{split}
 \DM_{\rm obs}(z) = & \DM_{\rm MW,ISM}+\DM_{\rm MW,halo} \\ 
 & +\DM_{\rm cos}(z) +\frac{\DM_{\rm host}}{1+z}.
\end{split}
\end{equation}
The first two terms arise from the Galactic interstellar medium (ISM) and halo. The rest-frame host contribution includes the host galaxy, its halo, and the local source environment. The cosmic contribution includes intergalactic gas and intervening halos. The mean cosmic DM is given by \citep{Deng2014ApJ,Zhou2014}
\begin{align}\label{eq:DMcosmic}
 \langle {{\rm DM_{cos}}(z_s)} \rangle &= \int^{z_s}_0\frac{cW(z){\rm d}z}{H(z)}, \\
 W(z) &= \frac{3\Omega_b H^2_0\chi_ef_{\rm d}(z)(1+z)}{8\pi G m_p},
\end{align}
where $H_0$ is the Hubble constant, $G$ is the gravitational constant, $\Omega_b$ is the baryon density parameter, $m_p$ is the proton mass, and $\chi_e=X_p+Y_p/2=0.88$ for primordial hydrogen and helium mass fractions $(X_p,Y_p)=(0.76,0.24)$. We use the TNG300-calibrated free-electron abundance $f_{\rm d}$ relative to fully ionized primordial baryons \citep{Takahashi2021}. For a flat Universe with the CPL parameterization $w=w_0 + w_az/(1+z)$ \citep{Chevallier2001}, 
\begin{equation}\label{Hz}
\begin{split}
H(z) = &H_0 \bigg[  \Omega_m (1+z)^3  \\ 
&+ {\rm \Omega_{DE}}(1+z)^{3(1+ w_0 + w_a)} \mathrm e^{-3 w_a z/(1+z)} \bigg]^{1/2}, 
\end{split}
\end{equation}
where $\Omega_m$ and $\Omega_{\rm DE}$ are the matter and DE density parameters, respectively. The parameters $w_0$ and $w_a$ describe the present-day EoS and its evolution. The $w$CDM model has $w_a=0$ and $w_0=w$, while $\Lambda$CDM has $(w_0,w_a)=(-1,0)$. 

For statistically isotropic electron fluctuations, we define the electron power spectrum as
\begin{equation}
\langle\widetilde\delta_e(\boldsymbol k,z)\widetilde\delta_e(\boldsymbol k',z)\rangle
=(2\pi)^3\delta_{\rm D}(\boldsymbol k+\boldsymbol k') P_{ee}(k,z),
\end{equation}
where a tilde denotes a Fourier transform, and $\delta_{\rm D}$ is the Dirac delta function. The angular DM spectrum in the Limber approximation is \citep{Takahashi2021,Reischke2023MNRAS}
\begin{equation}\label{eq:dm-angular-spectrum}
C_{\DM}(\ell,z_s)\simeq\int_0^{z_s}\frac{c\,\dd z}{H(z)}
\frac{W^2(z)}{r^2(z)}P_{ee}\!\left(\frac{\ell}{r(z)},z\right),
\end{equation}
where $r(z)=\int_0^z c\,\dd z'/H(z')$ is the comoving radial distance. The cosmic-DM variance is \citep{Reischke2023MNRAS}
\begin{equation}
\sigma_{\rm DM_{\rm cos}}^2(z_s)=\sum_\ell\frac{2\ell+1}{4\pi}C_{\DM}(\ell,z_s).
\end{equation}
In the flat-sky approximation, this becomes \citep{McQuinn2014,Takahashi2021}
\begin{equation}\label{eq:dm-variance}
\sigma_{\rm DM_{\rm cos}}^2(z_s)\simeq\frac{1}{2\pi}\int_0^{z_s}\frac{c\,\dd z}{H(z)}W^2(z)\int_0^\infty\dd k\,kP_{ee}(k,z).
\end{equation}
Wavenumbers and spectra in these integrals are expressed in consistent comoving units.

The electron spectrum is obtained by multiplying the TNG300-calibrated electron bias \citep{Takahashi2021} with the nonlinear matter power spectrum from the ten-parameter \texttt{GokuNEmu} emulator, which is based on dark-matter-only (DMO) simulations \citep{GokuEmu2025,Yang2026},
\begin{equation}
 P_{ee}(k, z \,| \boldsymbol\theta)
 =b_e^2(k,z)P_{\rm DMO}^{\rm Goku}(k, z \, |\boldsymbol\theta).
 \label{eq:electron-spectrum}
\end{equation}
Both spectra describe fractional density fluctuations. 
The background and matter power spectrum vary with cosmology, while $f_{\rm d}(z)$ and $b_e^2(k,z)$ retain their TNG300 calibration. Both \texttt{GokuNEmu} and the $b_e$ fit are used up to $k_{\max}=10\,h\,\mathrm{Mpc}^{-1}$. Previous studies have shown that gas cooling enhances electron clustering on smaller scales, and truncating the integral can underestimate the DM standard deviation by approximately 10\%--20\% \citep{Takahashi2021}. Therefore, we estimate this missing contribution using the median electron spectrum across 27 realizations with box sizes of $(50\,h^{-1}\,\mathrm{Mpc})^3$ from the second generation of CAMELS \citep{VillaescusaNavarro2021,Genel2026arXiv}. We define $I_K(z)=\int_0^K\dd k\,kP_{ee}(k,z)$, evaluated over the available CAMELS modes, and set $R_k(z)=I_{30}(z)/I_{10}(z)$, with the cutoffs given in $h,\mathrm{Mpc}^{-1}$. We multiply the inner integral in Eq.~\ref{eq:dm-variance}, truncated at $k_{\max}$, by $R_k(z)$ before integrating along the line of sight. For the reference cosmology, this correction increases $\sigma_{\rm DM_{\rm cos}}$ by approximately $4\%$--$8\%$ over $0<z_s\le2$ relative to the uncorrected prediction and yields a consistent value compared with the TNG300 sight-line dispersion.

\begin{figure*}[t]
 \centering
 \includegraphics[width=\textwidth]{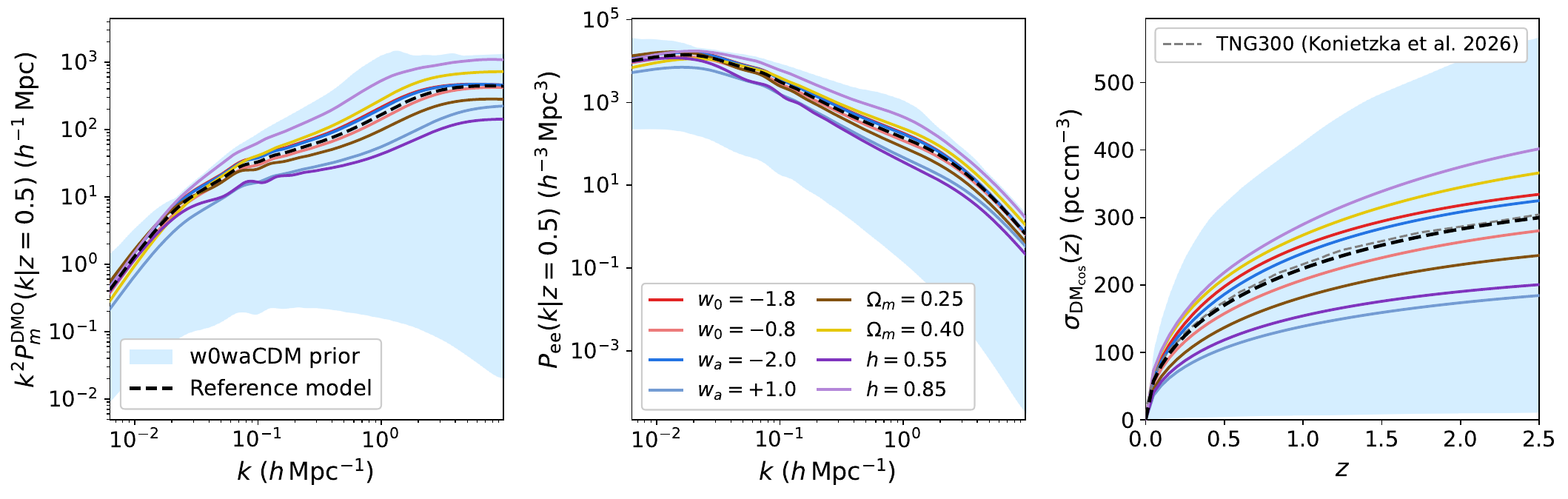}
\caption{Cosmological response of the matter power spectrum (left), electron spectrum (middle), and the cosmic-DM standard deviation (right). The spectra are shown at $z=0.5$. Black dashed curves show the TNG300 reference cosmology, and colored curves vary one parameter at a time. The blue regions indicate the envelope of sampled prior predictions of the $w_0w_a$CDM model. The gray dashed curve gives the standard deviation of 120,000 TNG300 sight lines \citep{Konietzka2026}.}
 \label{fig:power-response}
\end{figure*}

Figure~\ref{fig:power-response} illustrates how the cosmological response propagates from $P_{\rm DMO}$ to $P_{ee}$ and $\sigma_{\rm DM_{\rm cos}}$. For example, increasing $\Omega_m$ and increasing $w_0$ produce opposite responses in $\sigma_{\DM_{\rm cos}}$. The reference model adopts the same cosmology as TNG300 $(\Omega_m,h,\omega_b,w_0,w_a)=(0.3089,0.6774,0.022301,-1,0)$ \citep{Springel2018TNG}. Its predicted scatter is close to that of the TNG300 sight-line catalog \citep{Konietzka2026}. These responses demonstrate why cosmic-DM scatter should vary with cosmology.

\section{Analysis framework}

We adopt a lognormal distribution for $p_{\rm cos}$, whose parameters match the predicted mean and variance,
\begin{align}\label{eq:cosmic-lognormal}
 p_{\rm cos}(D|z) & = \frac{1}{\sqrt{2\pi}\,\sigma_{\rm cos}(z)\,D}
 \exp\bigg[-\frac{\left(\ln D-\mu_{\rm cos}(z)\right)^2} {2\sigma_{\rm cos}^2(z)}\bigg], \\
\mu_{\rm cos}(z) & =\ln \langle{\rm DM}_{\rm cos}(z)\rangle
 -\frac{\sigma_{\rm cos}^2(z)}{2}, \\
 \sigma_{\rm cos}^2(z) & =\ln\bigg[1+
 \frac{\sigma_{\rm cos}^2(z)}
      {\langle{\rm DM}_{\rm cos}(z)\rangle^2}\bigg],
\end{align}
where $D$ denotes the cosmic DM. We also adopt a lognormal rest-frame host distribution \citep{Macquart2020}
\begin{equation}\label{eq:host}
\begin{aligned}
p_{\rm {host}} &=  \frac{1}{\sqrt{2\pi}{\rm DM_{host}}\sigma_{\rm host}} {\rm exp} \left[ - \frac{({\rm ln \, DM}_{\rm host} - \mu_{\rm host})^2}{2\sigma_{\rm host}^2} \right].
\end{aligned}
\end{equation}
Since ${\rm DM_{host}^{obs}} = {\rm DM_{host}^{src}}/(1+z)$, the observer-frame PDF is
\begin{equation}
p_{\rm host}^{\rm obs}(D) = (1+z)p_{\rm host}^{\rm src}[(1+z)D].
\end{equation}
After subtracting the NE2025 Galactic ISM prediction \citep{Ocker2026} and a baseline Galactic halo contribution of $50\,\mathrm{pc\,cm^{-3}}$ \citep{Macquart2020,FRBData101038s4155002502566y}, the conditional probability of observing an FRB with extragalactic contribution $\DM_{\rm exgal}$ is 
\begin{align}
p(\DM_{\rm exgal},z|\boldsymbol{\theta}, \boldsymbol{\psi}) & = \int^{\DM_{\rm exgal}}_{0}{\rm d}\DM_{\rm cos}p_{\rm cos}(\DM_{\rm cos}, \, z|\boldsymbol{\theta}) \\
& \times p_{\rm host}^{\rm obs}(\DM_{\rm exgal}-\DM_{\rm cos},z\,|\boldsymbol{\psi}),
\end{align}
where $\boldsymbol{\theta}$ represents the cosmological parameters and $\boldsymbol{\psi}=(\mu_{\rm host},\,\sigma_{\rm host})$. For all localized FRBs, the total likelihood function can be written as 
\begin{equation}
 \mathcal L_{\rm FRB}=\prod_i^{N}p(\DM_{{\rm exgal},i},z_i \,|\boldsymbol{\theta}, \boldsymbol{\psi}),
\end{equation}
where the conditional probability has been normalized.

For Bayesian inference, we use \texttt{PyMultiNest} \citep{Buchner2014} as the sampler with 2000 live points. We adopt uniform priors with $\mu_{\rm host}\in[3,7]$, $\sigma_{\rm host}\in[0.3,1.5]$, $H_0\in[50,90]\,\mathrm{km\,s^{-1}\,Mpc^{-1}}$, $\Omega_m\in[0.1,0.5]$, $w_0\in[-3,0]$, and $w_a\in[-3,2]$. For the $w_0w_a$CDM model, we require $w_0+w_a<0$ so that DE decreases relative to matter toward high redshift. The physical baryon density has a BBN prior $\omega_b\sim\mathcal N(0.02233,0.00036^2)$ truncated at $5\sigma$ \citep{Pisanti2021}. For the FRB-only scenario, the primordial-spectrum and neutrino parameters are fixed at $A_s=2.1\times10^{-9}$, $n_s=0.9665$, $\alpha_s=0$, $\sum m_\nu=0.06\,\mathrm{eV}$, and $N_{\rm eff}=3.044$. Here, ``FRB-only'' denotes the FRB likelihood with these assumptions and the stated BBN prior, without CMB, BAO, or SNe Ia likelihoods.

For comparison and joint analysis, we use PantheonPlus relative SN distances with their full statistical and systematic covariance, without SH0ES calibration \citep{Brout2022}, and DESI DR2 BAO measurements with the released covariance \citep{DESI2025}. The CMB combination includes Planck 2018 high-$\ell$ TT/TE/EE, low-$\ell$ TT and EE power spectra \citep{Planck2020likelihood}, and the Planck PR4 lensing reconstruction \citep{Carron2022}. Likelihoods are computed using the cosmological Boltzmann code \texttt{CAMB} \citep{Lewis1999bs} and the cosmological inference framework \texttt{Cobaya} \citep{Torrado2021}.

\section{Results}\label{sec:results}
The current compilation contains 169 localized FRBs with confirmed host galaxies, listed in the Supplemental Material. We require $z>0.1$ and $\DM_{\rm exgal}>5\,\mathrm{pc\,cm^{-3}}$. At lower redshifts, peculiar velocities, local structures, and Galactic foregrounds can have substantial effects. Cosmic-DM fluctuations can approach the mean, and a two-moment PDF can miss the skewed tails associated with rare halo intersections. These effects motivate excluding the lowest-redshift sources from the baseline cosmological fit \citep{Sharma2025ApJ,Zhuge2026ApJ}, leaving 125 FRBs at $0.105\leq z\leq2.148$ as our sample.

\begin{figure}[t]
\centering
  \includegraphics[width=\linewidth]{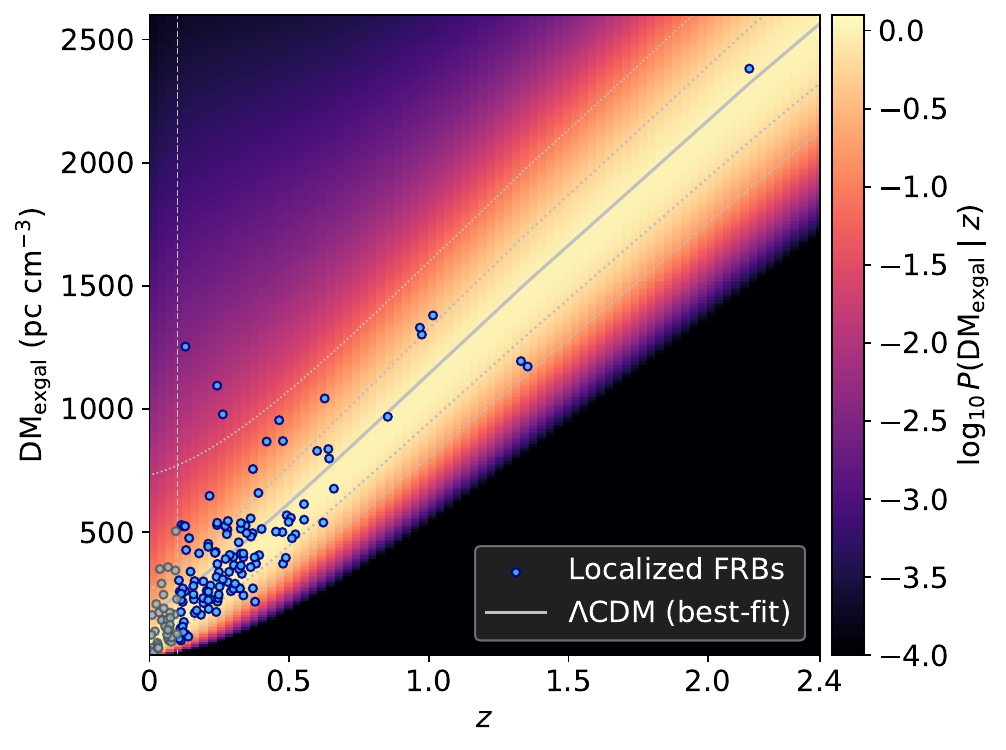}
  \includegraphics[width=\linewidth]{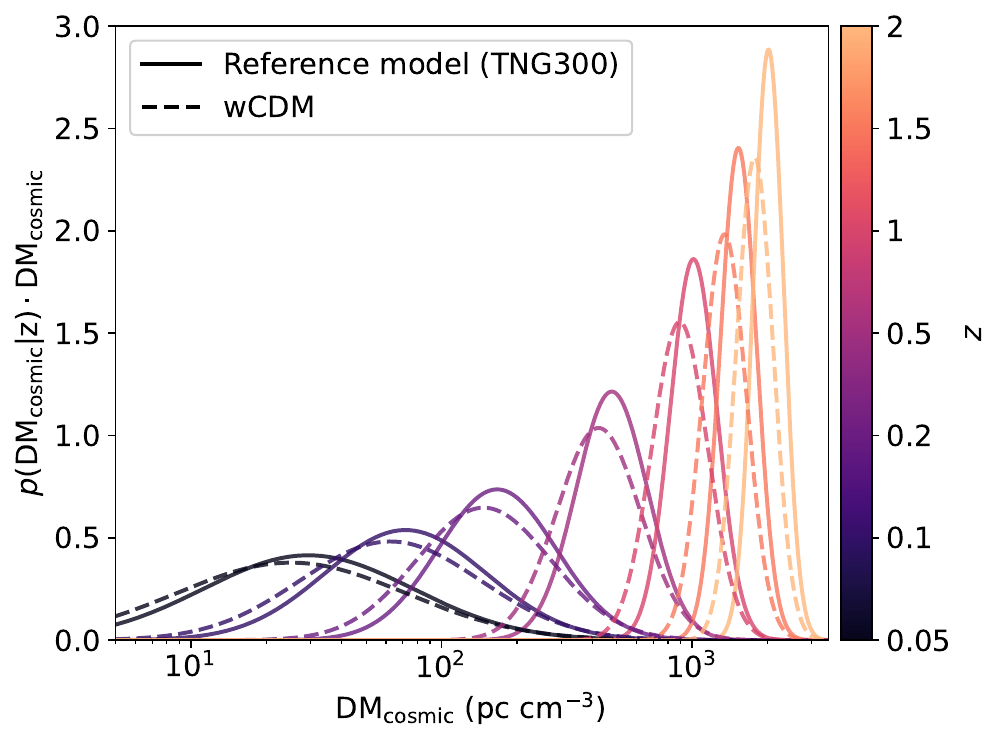}
\caption{The observed FRB $\DM_{\rm exgal}$–redshift relation and the distribution of $\DM_{\rm cos}$. The top panel shows localized FRBs with positive $\DM_{\rm exgal}$ and the reference $\Lambda$CDM prediction with  $(H_0,\Omega_m,\mu_{\rm host},\sigma_{\rm host})=(65.59,0.26,4.56,1.04)$. The solid curve marks the mean, and dotted curves enclose the central 68\% and 95\% intervals. Gray points at $z<0.1$ are excluded from the cosmological fit. The bottom panel shows $\DM_{\rm cos}p_{\rm cos}(\DM_{\rm cos}|z)$ for the reference cosmology (solid) and a $w$CDM example (dashed).}
 \label{fig:dm-distributions}
\end{figure}

The upper panel of Figure~\ref{fig:dm-distributions} compares the Galactic-subtracted observations with the convolution of the $\DM_{\rm cos}$ and $\DM_{\rm host}$ PDFs at the best-fit point of the $\Lambda$CDM model. The $p_{\rm exgal}$ distributions fit the observations well, and the baseline $\Lambda$CDM fit gives $H_0=66.49^{+6.99}_{-6.43}\,\mathrm{km\,s^{-1}\,Mpc^{-1}}$ and $\Omega_m=0.27^{+0.09}_{-0.07}$. In the bottom panel, we compare the TNG300 reference cosmology with a $w$CDM example having $(H_0,\Omega_m,w)=(72,0.33,-0.8)$. The cosmological response of the mean and variance shifts the peak and alters the fractional width and high-DM tail of $p_{\rm cos}$, illustrating the potential of localized FRBs to constrain cosmological parameters.

\begin{figure*}[!t]
 \centering
  \includegraphics[width=0.49\linewidth]{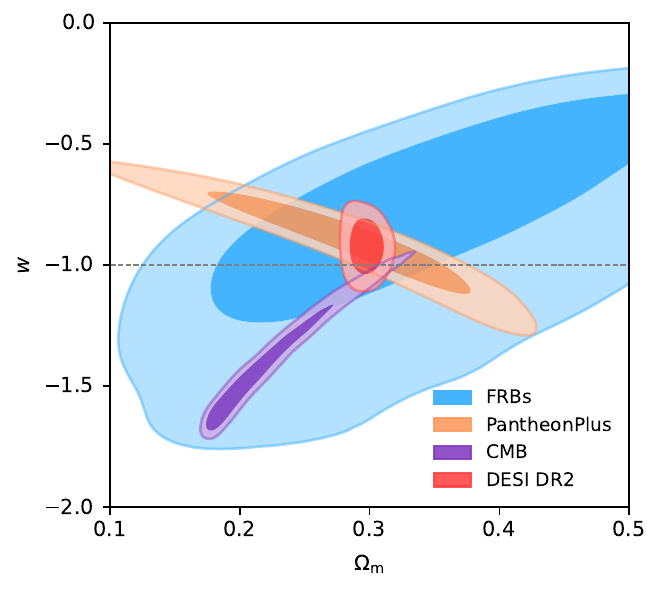}
  \includegraphics[width=0.49\linewidth]{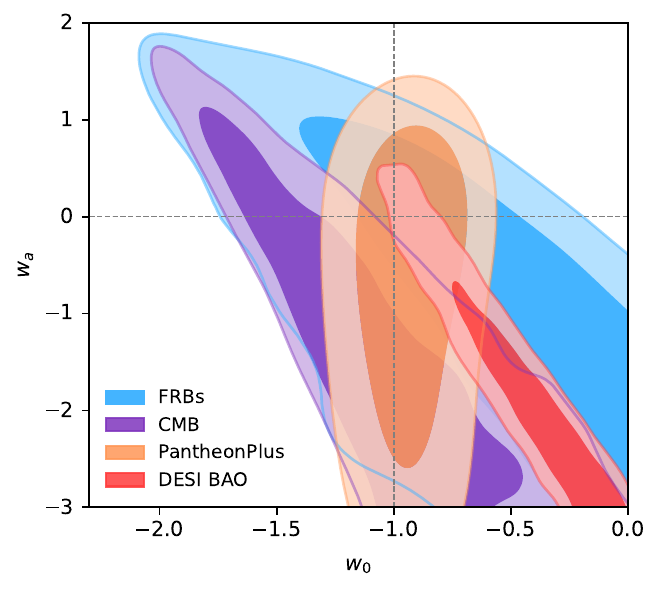}
\caption{The marginalized posterior constraints in the $\Omega_m-w$ plane for the $w$CDM model (left) and the $w_0-w_a$ plane for the $w_0w_a$CDM model (right). Gray dashed lines mark the $\Lambda$CDM framework. All contours show 68\% and 95\% credible intervals.}
 \label{fig:DE}
\end{figure*}

The constraints on the $w$CDM and $w_0w_a$CDM models are shown in Figure~\ref{fig:DE}, together with results from other cosmological observations. The full FRB-only posteriors appear in Appendix~\ref{app:frb-posteriors}. For the flat $w$CDM model, the FRB-only constraints are 
\begin{equation}
\left.
\begin{aligned}
& \Omega_m =0.33^{+0.10}_{-0.11}, &\\
&       w =-0.80^{+0.28}_{-0.38}, &
\end{aligned}
\right\}
\begin{gathered}
\mathrm{only \, \, FRBs}.
\end{gathered}
\end{equation}
These are the first simultaneous FRB constraints in the $\Omega_m$-$w$ plane without an external matter-density or DE measurement. The newly identified FRB direction is opposite to the strong SNe Ia anticorrelation, while BAO shows little global linear correlation. The corresponding sample correlation coefficients are $+0.54$, $-0.94$, and $-0.08$ for FRBs, SNe Ia, and DESI BAO, respectively. Near the reference cosmology, increasing $\Omega_m$ lowers $\langle\DM_{\rm cos}\rangle$ but raises $\sigma_{\DM_{\rm cos}}$, whereas increasing $w$ lowers both. The competition between these responses produces a degeneracy orientation opposite to that in the SNe Ia posterior. In Appendix~\ref{app:halo-effect}, we show that the inferred $w$ is stable against different halo prescriptions for the $z>0.1$ sample.

For flat $w_0w_a$CDM, the FRB-only constraints are 
\begin{equation}
\left.
\begin{aligned}
& \Omega_m =0.33\pm0.10, &\\
& w =-0.65^{+0.43}_{-0.57}, & \\
& w_a=-0.78^{+1.26}_{-1.25}, &
\end{aligned}
\right\}
\begin{gathered}
\mathrm{only \, \, FRBs}.
\end{gathered}
\end{equation}
The anticorrelation between $w_0$ and $w_a$ limits their separate determination. The $\Lambda$CDM point $(w_0,w_a)=(-1,0)$ lies within the FRB 68\% credible region. The models yield similar host-DM distributions, with $\mu_{\rm host}\sim4.4$ and $\sigma_{\rm host}\sim1.0$, broadly compatible with earlier estimates \citep{Sharma2026suppression}. 

Combining FRBs, Pantheon+, DESI DR2 BAO, and the CMB yields
\begin{equation}
\left.
\begin{aligned}
& \Omega_m = 0.305\pm0.005, &\\
& w =-0.986^{+0.022}_{-0.023}, &
\end{aligned}
\right\}
\begin{gathered}
\mathrm{FRBs+SNe \, Ia}\\
+\mathrm{BAO}+\mathrm{CMB},
\end{gathered}
\label{eq:joint-wcdm}
\end{equation}
in flat $w$CDM, with $w=-1$ inside the central 68\% interval. For the same data combination in flat $w_0w_a$CDM, the constraints are 
\begin{equation}
\left.
\begin{aligned}
& \Omega_m = 0.311\pm0.006, &\\
& w_0 = -0.846^{+0.054}_{-0.053}, & \\
& w_a = -0.548^{+0.192}_{-0.204}, &
\end{aligned}
\right\}
\begin{gathered}
\mathrm{FRBs+SNe \, Ia}\\
+\mathrm{BAO}+\mathrm{CMB},
\end{gathered}
\label{eq:joint-w0wacdm}
\end{equation}
which indicate a mild preference for dynamical DE over $\Lambda$CDM at approximately $2.6\sigma$, with a phantom-to-quintessence transition ($w_0>-1$ and $w_a<0$). 

\section{Discussion}\label{sec:discussion}
By propagating a common cosmology into the DM mean and variance, our emulator-based likelihood connects localized FRBs to both expansion and structure growth. It yields the first simultaneous FRB-only constraints on $\Omega_m$ and $w$, with a BBN prior and fixed primordial-spectrum and neutrino-sector parameters, and identifies a new positive degeneracy in these measurements. FRBs alone remain consistent with $\Lambda$CDM, whereas the full joint fit mildly favors evolution.

In our analysis, the emulator captures the cosmological response of nonlinear clustering beyond a calibration at one fiducial cosmology, but its finite training domain limits the present inference \citep{GokuEmu2025,Yang2026}. Although \texttt{GokuNEmu} covers a broad range of DE models, the broad FRB posteriors extend beyond this domain, where extrapolation accuracy remains unvalidated. Other emulators, such as EuclidEmulator2 \citep{Euclid2021MNRAS} and DarkEmulator2 \citep{Tanaka2026arXiv}, also have finite calibration ranges in parameters and scale. Additional simulations spanning the relevant joint parameter space, particularly $\Omega_m$, $H_0$, $w_0$, and $w_a$, are needed to extend the predictions and validate parameter recovery. 

The electron prescription introduces an additional limitation. The conversion from the matter spectrum to the electron spectrum relies on a calibrated $b_e^2(k,z)$, together with fixed $f_{\rm d}(z)$ and $R_k(z)$. Although TNG300 yields a range of predictions in good agreement with observations \citep{Nelson2018TNG, Springel2018TNG}, different subgrid prescriptions for stellar and AGN feedback can redistribute the gas and lead to a different small-scale suppression of the electron spectrum \citep{vanDaalen2020}. The constraints therefore remain conditional on this prescription and do not include its possible cosmological dependence or uncertainties in baryonic feedback. Matched hydrodynamic and DMO simulations varying both cosmology and feedback would calibrate these responses jointly. Simulation suites such as CAMELS provide a starting point for marginalizing over feedback uncertainty alongside host and cosmological parameters \citep{VillaescusaNavarro2021,Genel2026arXiv}.

The localized FRB sample will grow rapidly as additional facilities capable of precise localization come online and expand their survey footprints. CHIME/FRB Outriggers \citep{CHIME2026arXiv}, together with forthcoming instruments such as FAST Core Array \citep{FAST2024} and DSA-2000 \citep{DSA2000}, will greatly increase the number of bursts with positions suitable for host-galaxy identification. Combined with optical and infrared spectroscopic follow-up, these facilities should deliver much larger samples with measured redshifts over a broader redshift range. Such samples will sharpen measurements of the mean DM--redshift relation and the evolution of DM scatter, strengthening sensitivity to DE. With improved electron modeling and well-characterized detection and localization selection effects \citep{Sharma2025selection,WangYY2026}, these observational advances should drive substantial progress in FRB cosmology, enabling increasingly precise and complementary tests of cosmic expansion and structure growth.

\begin{acknowledgments}
This work is supported in part by NSFC under Grant No.~12588101, the Strategic Priority Research Program of the Chinese Academy of Sciences (grant No.~XDB0550400), the Postdoctoral Fellowship Program of CPSF (No.~GZB20250738), the China Postdoctoral Science Foundation (grant No.~2025M783236), and the Jiangsu Funding Program for Excellent Postdoctoral Talent (No.~2025ZB209).
\\ 
\end{acknowledgments}

\renewcommand{\bibsection}{}

\bibliographystyle{apsrev4-1}
\bibliography{reference.bib}

\appendix

\section{The effect of the Galactic halo contribution}\label{app:halo-effect}

\begin{figure}[!htbp]
 \centering
 \includegraphics[width=\columnwidth]{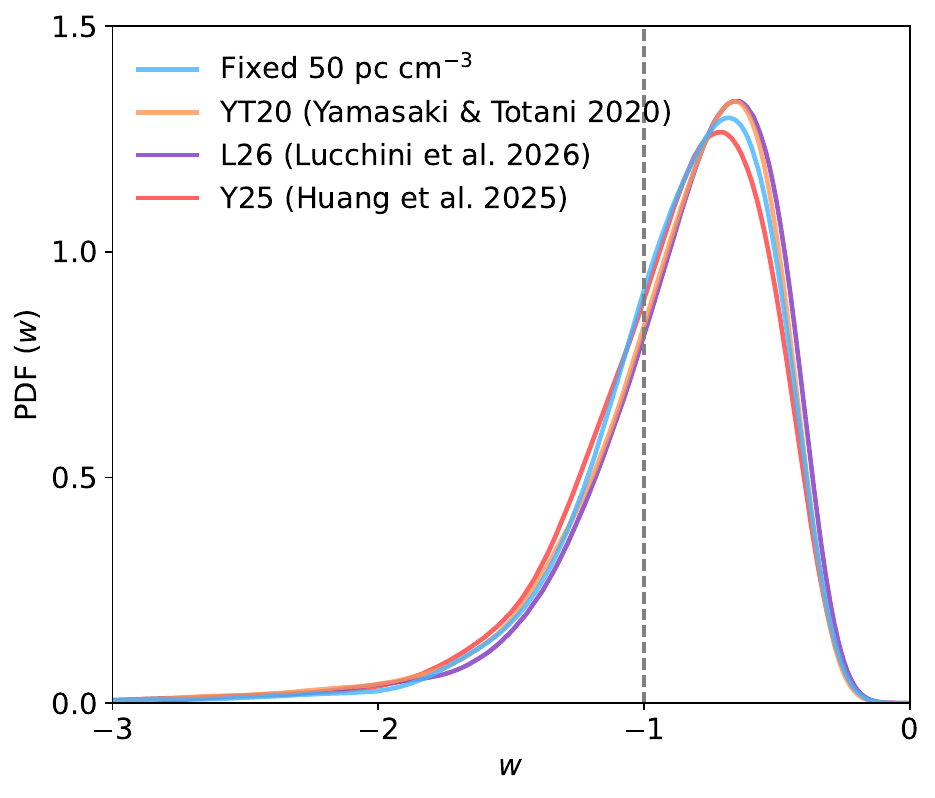}
\caption{FRB-only $w$CDM posteriors for Galactic halo prescriptions YT20, L26, and Y25, compared with the baseline assuming $50\,\mathrm{pc\,cm^{-3}}$. All fits use the same baseline sample. The dashed line marks the $\Lambda$CDM scenario.}
 \label{fig:halo-robustness}
\end{figure}

The results shown in the main text assume $\DM_{\rm MW,halo}=50\,\mathrm{pc\,cm^{-3}}$. We also test the dependence of our results on the Galactic foreground by replacing the fixed halo contribution with three direction-dependent predictions. YT20 includes spherical and disk-like hot-gas components \citep{HaloYT2020}, while L26 uses the ENGAWA simulations and accounts for halo anisotropy \citep{HaloLucchini2026}. Y25 is based on a model of the local Universe within $D<120\,\mathrm{Mpc}$ that includes Local Group galaxies such as the Large and Small Magellanic Clouds and the Triangulum Galaxy \citep{HaloHuang2025}. Figure~\ref{fig:halo-robustness} shows the marginalized $w$ posteriors, using the same sample and likelihood settings. The median $w$ values are $-0.79$, $-0.77$, and $-0.82$ for YT20, L26, and Y25, respectively, compared with $-0.804$ for the baseline. The inferred $w$ is stable against these halo prescriptions for the $z>0.1$ sample.

\clearpage
\onecolumngrid
\section{FRB-only parameter posteriors}\label{app:frb-posteriors}

\begin{figure*}[!htbp]
\centering
\includegraphics[width=\textwidth,height=0.76\textheight,keepaspectratio]{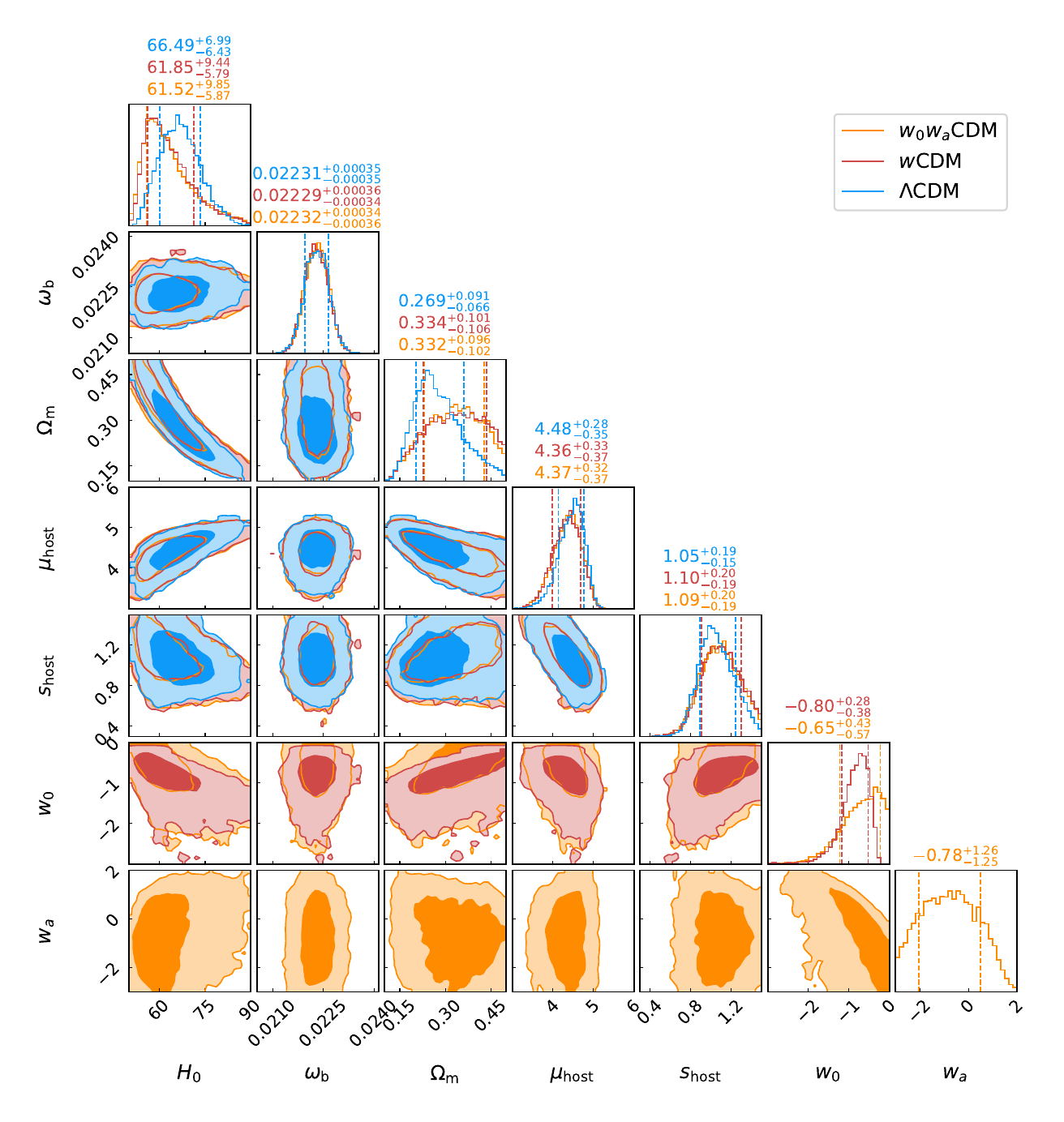}
\caption{Posterior distributions for flat $\Lambda$CDM (blue), $w$CDM (red), and $w_0w_a$CDM (orange) using FRB data alone. The contours represent the 68\% and 99\% credible intervals, respectively. The values and dashed lines in the marginal distributions represent the 68\% credible intervals.}
 \label{fig:frb-full-posteriors}
\end{figure*}

Figure~\ref{fig:frb-full-posteriors} shows the cosmological and host-population posteriors in flat $\Lambda$CDM, $w$CDM, and $w_0w_a$CDM. The joint distributions show the correlations with the host population and the broader cosmological posteriors when DE parameters are freed. The narrow $\omega_b$ posterior largely reflects the BBN prior and should not be interpreted as an independent FRB baryon-density measurement.

\end{document}


\title{Localized Fast Radio Bursts catalog used in ``First $\Omega_{\rm m}-w$ Measurement Using Only Fast Radio Burst Observations"}

\maketitle

Table~\ref{tab:frb-catalog} lists the 169 retained localized FRBs with adopted host redshifts. Of these, 125 pass the baseline redshift and DM cuts and span $0.105\leq z\leq2.148$. Of the 44 sources below $z=0.1$, 41 pass the DM cut and form the low-redshift comparison sample. The Galactic halo predictions are tabulated for comparison.

We exclude FRB20220330D and FRB20221027A because each has two plausible host galaxies at substantially different spectroscopic redshifts. We also exclude FRB20220208A, FRB20230216A, FRB20230808F, FRB20231010A, and FRB20231210F because their low host-association posteriors leave the adopted redshifts uncertain [Refs.~(27), (29), and (41)]. The table notes describe cases retained despite host-association uncertainty.

\begingroup
\small
\setlength{\tabcolsep}{3pt}
\renewcommand{\arraystretch}{1.08}
\setlength{\LTpre}{6pt}
\setlength{\LTpost}{8pt}
\begin{longtable}{@{}lrrrrrrrrl@{}}
\caption{Localized FRBs and Galactic foreground predictions. All DM values are in $\mathrm{pc\,cm^{-3}}$.}\label{tab:frb-catalog}\\
\hline\hline
Name & Redshift & $\DM_{\rm obs}$ & R.A. & Decl. & $\DM_{\rm MW,ISM}$ & \multicolumn{3}{c}{$\DM_{\rm MW,halo}$} & Ref. \\
 & & $(\mathrm{pc\,cm^{-3}})$ & (deg, J2000) & (deg, J2000) & NE2025 & YT20 & L26 & Y25 & \\
\hline
\endfirsthead
\caption[]{Localized FRBs and Galactic foreground predictions (continued).}\\
\hline\hline
Name & Redshift & $\DM_{\rm obs}$ & R.A. & Decl. & $\DM_{\rm MW,ISM}$ & \multicolumn{3}{c}{$\DM_{\rm MW,halo}$\textsuperscript{f}} & Ref. \\
 & & $(\mathrm{pc\,cm^{-3}})$ & (deg, J2000) & (deg, J2000) & NE2025\textsuperscript{e} & YT20 & L26 & Y25 & \\
\hline
\endhead
\hline
\multicolumn{10}{r}{\footnotesize Continued on the next page}\\
\endfoot
\hline\hline
\endlastfoot
FRB20121102A & 0.1927 & 557.0 & 82.9946 & 33.1479 & 207.9 & 40.9 & 83.8 & 23.6 & (1,2,3) \\
FRB20171020A\textsuperscript{a} & 0.00867 & 114.0 & 333.7500 & -19.6667 & 34.4 & 39.5 & 14.2 & 27.1 & (4) \\
FRB20180301A & 0.3304 & 536.0 & 93.2270 & 4.6710 & 156.2 & 38.1 & 56.8 & 24.8 & (5,3) \\
FRB20180814A\textsuperscript{a} & 0.068 & 189.4 & 65.6833 & 73.6644 & 79.3 & 35.7 & 40.6 & 42.7 & (6) \\
FRB20180916B\textsuperscript{a} & 0.0337 & 348.76 & 29.5033 & 65.7168 & 125.6 & 43.1 & 70.4 & 24.9 & (2,3) \\
FRB20180924B & 0.3214 & 362.16 & 326.1054 & -40.9000 & 39.8 & 45.5 & 16.2 & 30.3 & (2,7,3) \\
FRB20181030A\textsuperscript{a} & 0.00385 & 103.5 & 158.5838 & 73.7514 & 37.1 & 31.6 & 32.4 & 43.3 & (8) \\
FRB20181112A & 0.4755 & 589.0 & 327.3483 & -52.9709 & 39.5 & 45.0 & 17.2 & 37.5 & (2,7) \\
FRB20181220A\textsuperscript{a} & 0.0275 & 209.4 & 348.7167 & 48.3403 & 109.3 & 42.8 & 40.8 & 50.3 & (9) \\
FRB20181223C\textsuperscript{a} & 0.03024 & 112.5 & 180.9292 & 27.5525 & 21.6 & 30.9 & 10.5 & 34.4 & (9) \\
FRB20190102C & 0.2913 & 364.545 & 322.4158 & -79.4757 & 55.0 & 47.0 & 25.6 & 68.1 & (2,7,3) \\
FRB20190110C & 0.12244 & 221.6 & 249.3180 & 41.4430 & 37.0 & 38.9 & 21.0 & 66.4 & (10) \\
FRB20190303B\textsuperscript{a,d} & 0.0656 & 193.43 & 128.6877 & 66.0059 & 41.0 & 31.3 & 37.3 & 45.3 & (11) \\
FRB20190417A & 0.12817 & 1379.2 & 294.7745 & 59.3269 & 76.2 & 43.3 & 46.0 & 39.8 & (12) \\
FRB20190418A\textsuperscript{a} & 0.0713 & 184.5 & 65.8167 & 16.0672 & 59.4 & 32.3 & 44.2 & 28.9 & (9) \\
FRB20190425A\textsuperscript{a} & 0.0312 & 128.2 & 255.6750 & 21.5764 & 48.3 & 49.0 & 25.2 & 48.7 & (9) \\
FRB20190502A\textsuperscript{d} & 0.2803 & 625.74 & 165.0863 & 59.9156 & 31.1 & 30.4 & 25.0 & 36.7 & (11) \\
FRB20190520B & 0.2418 & 1204.7 & 240.5179 & -11.2881 & 60.4 & 69.2 & 25.3 & 49.5 & (13,14,3) \\
FRB20190523A & 0.66 & 760.8 & 207.0650 & 72.4697 & 35.0 & 32.4 & 27.6 & 47.7 & (2,15) \\
FRB20190608B & 0.1178 & 339.5 & 334.0200 & -7.8982 & 36.1 & 38.9 & 16.8 & 29.9 & (2,7,3) \\
FRB20190611B & 0.3778 & 322.2 & 320.7454 & -79.3976 & 55.4 & 47.2 & 25.8 & 64.6 & (2,7,3) \\
FRB20190614D & 0.6\textsuperscript{b} & 959.0 & 65.0754 & 73.7067 & 79.8 & 35.7 & 40.6 & 42.7 & (16) \\
FRB20190711A & 0.522 & 594.6 & 329.4196 & -80.3580 & 54.0 & 46.2 & 25.4 & 92.3 & (2,7) \\
FRB20190714A & 0.2365 & 504.7 & 183.9797 & -13.0210 & 35.7 & 36.4 & 18.5 & 39.5 & (2,7,3) \\
FRB20191001A & 0.234 & 506.92 & 323.3517 & -54.7483 & 41.8 & 46.7 & 18.3 & 35.4 & (2,7,3) \\
FRB20191106C & 0.10775 & 332.2 & 199.5801 & 42.9997 & 24.6 & 31.7 & 11.0 & 43.6 & (10) \\
FRB20191228A & 0.2432 & 297.5 & 344.4304 & -29.5941 & 30.2 & 36.9 & 9.9 & 39.8 & (5,7) \\
FRB20200120E\textsuperscript{a} & 0.0008 & 87.77 & 149.4862 & 68.8256 & 36.3 & 31.1 & 32.1 & 40.5 & (17,18,19) \\
FRB20200223B\textsuperscript{a} & 0.06024 & 201.8 & 8.2695 & 28.8313 & 42.8 & 33.6 & 37.2 & 48.2 & (10) \\
FRB20200430A & 0.1608 & 380.1 & 229.7064 & 12.3763 & 29.5 & 42.2 & 14.5 & 40.7 & (2,7,3) \\
FRB20200723B\textsuperscript{a} & 0.0085 & 243.99 & 190.1583 & -5.1350 & 31.4 & 35.6 & 13.8 & 34.7 & (20) \\
FRB20200906A & 0.3688 & 577.8 & 53.4963 & -14.0832 & 32.7 & 30.2 & 27.5 & 36.0 & (5,7,3) \\
FRB20201123A\textsuperscript{a} & 0.0507 & 433.55 & 263.6692 & -50.7672 & 193.9 & 117.0 & 45.9 & 116.9 & (21) \\
FRB20201124A\textsuperscript{a} & 0.0979 & 413.52 & 77.0146 & 26.0607 & 136.1 & 36.2 & 47.1 & 38.4 & (22,3) \\
FRB20210117A & 0.2145 & 729.1 & 339.9792 & -16.1514 & 32.3 & 37.2 & 12.6 & 33.2 & (3,7,23) \\
FRB20210320C & 0.2796 & 384.8 & 204.4588 & -16.1227 & 37.9 & 42.1 & 17.4 & 30.8 & (3,7) \\
FRB20210405I\textsuperscript{a} & 0.0656 & 565.17 & 255.3396 & -49.5452 & 444.2 & 144.0 & 48.2 & 101.6 & (24) \\
FRB20210410D & 0.1415 & 578.78 & 326.0863 & -79.3182 & 53.9 & 46.5 & 24.9 & 76.4 & (3,25) \\
FRB20210603A & 0.1772 & 500.147 & 10.2741 & 21.2263 & 36.9 & 32.4 & 29.7 & 51.9 & (26) \\
FRB20210807D & 0.1293 & 251.9 & 299.2214 & -0.7624 & 117.1 & 73.6 & 49.0 & 77.6 & (3,7) \\
FRB20211127I\textsuperscript{a} & 0.0469 & 234.83 & 199.8088 & -18.8379 & 40.3 & 41.6 & 18.3 & 35.6 & (3,7) \\
FRB20211203C & 0.3437 & 636.2 & 204.5625 & -31.3803 & 60.5 & 49.8 & 29.7 & 52.5 & (3,7) \\
FRB20211212A\textsuperscript{a} & 0.0707 & 206.0 & 157.3508 & 1.3604 & 34.7 & 32.2 & 25.4 & 41.7 & (3,7) \\
FRB20220105A & 0.2785 & 583.0 & 208.8034 & 22.4662 & 23.4 & 34.4 & 8.3 & 26.1 & (3,7) \\
FRB20220204A & 0.4012 & 612.2 & 274.2250 & 69.7225 & 50.0 & 37.0 & 42.6 & 50.4 & (27,18) \\
FRB20220207C\textsuperscript{a} & 0.0433 & 262.38 & 310.1996 & 72.8823 & 71.0 & 39.2 & 45.0 & 47.7 & (28,27,18) \\
FRB20220222C & 0.853 & 1071.2 & 203.9045 & -28.0269 & 53.2 & 47.4 & 24.8 & 68.8 & (29) \\
FRB20220224C & 0.6271 & 1140.2 & 166.6775 & -22.9399 & 47.8 & 37.7 & 33.1 & 53.4 & (29) \\
FRB20220307B & 0.2481 & 499.27 & 350.8746 & 72.1924 & 111.8 & 41.3 & 41.3 & 44.5 & (28,27,18) \\
FRB20220310F & 0.477958 & 462.24 & 134.7204 & 73.4908 & 41.2 & 31.8 & 37.4 & 50.2 & (28,27,18) \\
FRB20220319D\textsuperscript{a} & 0.0111 & 110.98 & 32.1779 & 71.0350 & 99.4 & 39.7 & 44.3 & 39.4 & (28,27,18,30) \\
FRB20220418A & 0.622 & 623.25 & 219.1054 & 70.0959 & 34.9 & 32.8 & 26.5 & 41.4 & (28,27,18) \\
FRB20220501C & 0.381 & 449.5 & 352.3792 & -32.4907 & 28.3 & 35.3 & 9.6 & 39.7 & (7) \\
FRB20220506D & 0.3004 & 396.97 & 318.0450 & 72.8273 & 78.1 & 39.8 & 43.3 & 48.9 & (28,27,18) \\
FRB20220509G\textsuperscript{a} & 0.0894 & 269.53 & 282.6700 & 70.2438 & 54.4 & 37.7 & 45.4 & 53.2 & (28,27,18) \\
FRB20220529A & 0.1839 & 250.2 & 19.1042 & 20.6324 & 36.6 & 31.7 & 30.7 & 43.4 & (31) \\
FRB20220610A & 1.015 & 1458.1 & 351.0733 & -33.5138 & 28.6 & 35.6 & 10.0 & 41.7 & (32,7) \\
FRB20220717A & 0.36295 & 637.34 & 293.3042 & -19.2877 & 106.1 & 86.5 & 48.5 & 87.5 & (33) \\
FRB20220725A & 0.1926 & 290.4 & 353.3167 & -35.9903 & 28.5 & 35.4 & 11.4 & 44.9 & (7) \\
FRB20220726A & 0.3619 & 686.55 & 73.9458 & 69.9291 & 81.2 & 35.3 & 40.0 & 45.8 & (27,18) \\
FRB20220801A & 0.37 & 413.0 & 54.9635 & 70.1816 & 92.0 & 37.4 & 39.6 & 37.0 & (34) \\
FRB20220825A & 0.2414 & 651.24 & 311.9813 & 72.5850 & 73.1 & 39.5 & 44.5 & 41.1 & (28,27,18) \\
FRB20220831A & 0.262 & 1146.25 & 338.6958 & 70.5388 & 118.2 & 42.1 & 41.2 & 45.3 & (18) \\
FRB20220912A\textsuperscript{a} & 0.0771 & 219.46 & 347.2704 & 48.7066 & 115.7 & 43.4 & 41.2 & 33.7 & (35) \\
FRB20220914A & 0.1139 & 631.28 & 282.0567 & 73.3369 & 52.9 & 37.0 & 45.0 & 50.7 & (28,27,18) \\
FRB20220918A & 0.491 & 656.8 & 17.5917 & -70.8114 & 39.0 & 38.7 & 18.2 & 78.7 & (7) \\
FRB20220920A & 0.1585 & 314.99 & 240.2571 & 70.9188 & 38.7 & 34.1 & 30.2 & 49.2 & (28,27,18) \\
FRB20221012A & 0.2847 & 441.08 & 280.7987 & 70.5242 & 53.2 & 37.5 & 44.9 & 48.8 & (28,27) \\
FRB20221022A\textsuperscript{a} & 0.0149 & 116.843 & 48.6292 & 86.8719 & 54.8 & 34.7 & 45.5 & 43.3 & (36) \\
FRB20221029A & 0.975 & 1391.05 & 141.9625 & 72.4523 & 39.2 & 31.6 & 35.2 & 45.2 & (27,18) \\
FRB20221101B & 0.2395 & 490.7 & 342.2167 & 70.6812 & 123.1 & 42.1 & 41.7 & 44.1 & (27,18) \\
FRB20221106A & 0.2044 & 343.8 & 56.7042 & -25.5698 & 32.6 & 30.7 & 25.5 & 35.6 & (7) \\
FRB20221113A & 0.2505 & 411.4 & 71.4125 & 70.3074 & 83.4 & 35.5 & 39.8 & 48.2 & (27,18) \\
FRB20221116A & 0.2764 & 640.6 & 21.2083 & 72.6539 & 99.5 & 39.8 & 42.2 & 45.1 & (27,18) \\
FRB20221203A & 0.51 & 602.2 & 315.1295 & 72.0376 & 77.4 & 39.9 & 43.6 & 45.6 & (34) \\
FRB20221219A & 0.553 & 706.708 & 257.6292 & 71.6268 & 43.5 & 35.3 & 36.1 & 51.2 & (27,18) \\
FRB20230124A\textsuperscript{a} & 0.0939 & 590.6 & 231.9167 & 70.9681 & 37.1 & 33.6 & 28.6 & 43.5 & (27,18) \\
FRB20230125D & 0.3265 & 640.08 & 150.2050 & -31.5447 & 76.9 & 41.8 & 46.5 & 51.4 & (29) \\
FRB20230203A & 0.1464 & 420.1 & 151.6616 & 35.6941 & 31.0 & 29.9 & 24.9 & 29.5 & (37) \\
FRB20230222A & 0.1223 & 706.1 & 106.9604 & 11.2245 & 133.5 & 37.0 & 46.4 & 32.7 & (37) \\
FRB20230222B & 0.11 & 187.8 & 238.7391 & 30.8987 & 29.9 & 39.0 & 16.1 & 54.2 & (37) \\
FRB20230307A & 0.2706 & 608.9 & 177.7833 & 71.6956 & 34.4 & 31.5 & 28.6 & 42.6 & (27,18) \\
FRB20230311A\textsuperscript{c} & 0.1918 & 364.3 & 91.1097 & 55.9460 & 82.9 & 34.3 & 39.1 & 31.2 & (37) \\
FRB20230501A & 0.3015 & 532.5 & 340.0292 & 70.9222 & 117.2 & 41.9 & 41.1 & 44.1 & (27,18) \\
FRB20230503E & 0.32\textsuperscript{b} & 483.74 & 238.4300 & -83.7753 & 84.3 & 53.7 & 44.5 & 78.0 & (29) \\
FRB20230506C & 0.3896 & 772.0 & 12.0998 & 42.0061 & 63.6 & 36.0 & 45.1 & 41.4 & (38) \\
FRB20230521B & 1.354 & 1342.9 & 351.0360 & 71.1380 & 120.8 & 41.9 & 42.9 & 44.0 & (18) \\
FRB20230526A & 0.157 & 361.4 & 22.2333 & -52.7173 & 30.4 & 34.1 & 12.0 & 48.3 & (7) \\
FRB20230613A & 0.3923 & 483.51 & 356.8527 & -27.0528 & 27.6 & 34.0 & 7.9 & 43.5 & (29) \\
FRB20230626A & 0.327 & 451.2 & 235.6292 & 71.1335 & 37.8 & 33.8 & 29.3 & 44.2 & (27,18) \\
FRB20230628A & 0.127 & 345.15 & 166.7875 & 72.2818 & 35.5 & 31.4 & 30.4 & 44.8 & (27,18) \\
FRB20230703A & 0.1184 & 291.3 & 184.6245 & 48.7299 & 25.5 & 30.7 & 14.6 & 40.3 & (37) \\
FRB20230708A & 0.105 & 411.51 & 303.1167 & -55.3563 & 57.8 & 58.0 & 22.3 & 43.6 & (7,39,40) \\
FRB20230712A & 0.4525 & 586.96 & 167.3583 & 72.5578 & 35.6 & 31.5 & 30.4 & 44.8 & (27,18) \\
FRB20230718A\textsuperscript{a} & 0.0357 & 477.0 & 128.1625 & -40.4519 & 490.2 & 56.4 & 90.9 & 31.1 & (7) \\
FRB20230730A & 0.2115 & 312.5 & 54.6645 & 33.1593 & 75.3 & 33.8 & 40.5 & 28.3 & (37) \\
FRB20230814A & 0.5535 & 696.4 & 335.9748 & 73.0259 & 96.6 & 40.5 & 40.5 & 37.9 & (18) \\
FRB20230902A & 0.3619 & 440.1 & 52.1417 & -47.3335 & 32.8 & 32.6 & 20.3 & 43.4 & (7) \\
FRB20230907D & 0.4638 & 1030.79 & 187.1425 & 8.6581 & 26.9 & 33.0 & 9.7 & 38.6 & (29) \\
FRB20230913G & 0.3 & 518.7 & 305.0372 & 70.7928 & 70.2 & 39.6 & 45.5 & 50.8 & (34) \\
FRB20230926A\textsuperscript{a} & 0.0553 & 222.8 & 269.1249 & 41.8143 & 55.6 & 44.2 & 39.6 & 59.6 & (37) \\
FRB20230930A\textsuperscript{a} & 0.0925 & 456.0 & 10.5072 & 41.4173 & 62.2 & 36.0 & 45.5 & 35.6 & (38) \\
FRB20231005A\textsuperscript{a} & 0.0713 & 189.4 & 246.0280 & 35.4487 & 34.3 & 39.6 & 18.8 & 48.5 & (37) \\
FRB20231011A\textsuperscript{a} & 0.0783 & 186.3 & 18.2411 & 41.7491 & 63.5 & 35.4 & 44.8 & 28.0 & (37) \\
FRB20231017A & 0.245 & 344.2 & 346.7543 & 36.6527 & 62.4 & 39.1 & 46.9 & 47.3 & (37) \\
FRB20231020B & 0.4775 & 952.2 & 57.2782 & -37.7699 & 32.9 & 31.7 & 23.2 & 41.3 & (29) \\
FRB20231025B & 0.3238 & 368.7 & 270.7881 & 63.9891 & 49.0 & 37.8 & 40.5 & 48.0 & (37) \\
FRB20231120A\textsuperscript{a} & 0.0368 & 438.9 & 143.9833 & 73.2847 & 39.1 & 31.6 & 35.1 & 48.8 & (27,18) \\
FRB20231123A\textsuperscript{a} & 0.0729 & 302.1 & 82.6232 & 4.4755 & 79.6 & 34.2 & 38.8 & 30.8 & (37) \\
FRB20231123B & 0.2621 & 396.7 & 242.5375 & 70.7851 & 39.1 & 34.3 & 30.7 & 52.1 & (27,18) \\
FRB20231201A & 0.1119 & 169.4 & 54.5893 & 26.8177 & 60.0 & 32.5 & 43.8 & 24.9 & (37) \\
FRB20231204A\textsuperscript{a} & 0.0644 & 221.0 & 207.9990 & 48.1160 & 27.4 & 32.2 & 13.0 & 38.0 & (37,6) \\
FRB20231206A\textsuperscript{a} & 0.0659 & 457.7 & 112.4428 & 56.2563 & 50.4 & 31.8 & 43.2 & 37.7 & (37) \\
FRB20231220A & 0.3355 & 491.2 & 123.9087 & 73.6599 & 44.2 & 32.1 & 40.3 & 48.9 & (18) \\
FRB20231223C & 0.1059 & 165.8 & 259.5447 & 29.4979 & 48.8 & 46.3 & 28.6 & 54.3 & (37) \\
FRB20231226A & 0.1569 & 329.9 & 155.3625 & 6.1102 & 33.8 & 31.6 & 25.4 & 45.9 & (7) \\
FRB20231229A\textsuperscript{a} & 0.019 & 198.5 & 26.4678 & 35.1129 & 51.8 & 33.3 & 44.5 & 31.6 & (37) \\
FRB20231230A\textsuperscript{a} & 0.0298 & 131.4 & 72.7976 & 2.3940 & 53.1 & 32.1 & 44.0 & 30.9 & (37) \\
FRB20231230D & 0.506 & 676.5 & 118.7608 & 8.5141 & 68.9 & 34.3 & 41.0 & 40.3 & (34) \\
FRB20240104A & 1.33 & 1351.0 & 348.8740 & 72.8206 & 107.0 & 40.9 & 40.6 & 44.5 & (34) \\
FRB20240114A & 0.1306 & 527.7 & 321.9160 & 4.3294 & 51.1 & 45.0 & 30.4 & 43.3 & (42,43,44) \\
FRB20240117B & 0.643 & 881.0 & 53.8775 & -15.8523 & 32.5 & 30.2 & 27.0 & 38.4 & (34) \\
FRB20240119A & 0.376 & 483.1 & 224.4672 & 71.6118 & 36.2 & 33.2 & 28.0 & 44.2 & (18) \\
FRB20240123A & 0.968 & 1462.0 & 68.2625 & 71.9453 & 82.1 & 35.6 & 40.1 & 41.9 & (18) \\
FRB20240201A\textsuperscript{a} & 0.042729 & 374.5 & 149.9042 & 14.0880 & 33.8 & 30.9 & 27.3 & 54.7 & (7) \\
FRB20240203C & 0.2426 & 454.0 & 8.8975 & 15.8409 & 34.1 & 32.1 & 25.8 & 49.7 & (34) \\
FRB20240203D\textsuperscript{a} & 0.074 & 272.6 & 312.6191 & 73.9000 & 70.9 & 39.0 & 44.9 & 44.3 & (34) \\
FRB20240208A & 0.2385 & 260.2 & 159.2292 & -0.9544 & 35.0 & 32.5 & 24.9 & 41.0 & (45) \\
FRB20240209A & 0.1384 & 176.49 & 289.8875 & 86.0644 & 51.4 & 34.9 & 44.8 & 46.1 & (46,47) \\
FRB20240210A\textsuperscript{a} & 0.023686 & 283.73 & 8.7792 & -28.2707 & 26.4 & 32.4 & 8.2 & 54.8 & (7) \\
FRB20240213A & 0.1185 & 357.4 & 166.1683 & 74.0754 & 36.5 & 31.6 & 31.5 & 47.4 & (18) \\
FRB20240215A & 0.21 & 549.5 & 268.4413 & 70.2324 & 47.3 & 36.4 & 40.2 & 46.3 & (18) \\
FRB20240224A & 0.37 & 881.2 & 70.1149 & 73.5128 & 75.6 & 35.3 & 41.3 & 37.9 & (34) \\
FRB20240229A & 0.287 & 491.15 & 169.9835 & 70.6762 & 34.5 & 31.3 & 29.1 & 45.7 & (18) \\
FRB20240304A & 0.2423 & 652.6 & 136.3292 & -16.1666 & 65.1 & 37.1 & 45.3 & 44.5 & (7) \\
FRB20240304B & 2.148 & 2458.2 & 182.9970 & 11.8130 & 26.3 & 32.3 & 10.4 & 31.3 & (48) \\
FRB20240310A & 0.127 & 601.8 & 17.6219 & -44.4394 & 28.3 & 33.2 & 9.0 & 41.3 & (7) \\
FRB20240312D\textsuperscript{a} & 0.0495 & 330.0 & 51.6234 & -54.6010 & 34.4 & 33.8 & 20.5 & 44.9 & (34) \\
FRB20240318A & 0.112 & 256.4 & 150.3932 & 37.6164 & 31.4 & 29.9 & 26.0 & 31.8 & (7,49) \\
FRB20240525A & 0.3266 & 491.6 & 6.7829 & -6.8896 & 29.5 & 31.7 & 12.4 & 41.2 & (34) \\
FRB20240615B\textsuperscript{a} & 0.073 & 201.0 & 33.9636 & -14.6245 & 28.1 & 30.1 & 15.3 & 40.2 & (34) \\
FRB20240807A & 0.2098 & 518.1 & 23.3032 & -6.9842 & 28.2 & 30.4 & 15.0 & 39.6 & (50) \\
FRB20241014B & 0.3355 & 497.66 & 49.8391 & -58.3210 & 35.3 & 34.5 & 20.3 & 42.2 & (50) \\
FRB20241027B & 0.336 & 320.0 & 36.0290 & -20.7035 & 27.9 & 30.2 & 15.1 & 33.2 & (34) \\
FRB20241228A & 0.1614 & 246.53 & 216.3858 & 12.0250 & 25.7 & 38.0 & 9.6 & 30.5 & (51) \\
FRB20250106B & 0.3475 & 578.01 & 54.2957 & -16.2045 & 32.6 & 30.2 & 27.1 & 39.0 & (50) \\
FRB20250206A & 0.118 & 207.1 & 338.6708 & 12.1792 & 41.1 & 37.7 & 26.0 & 38.7 & (52) \\
FRB20250213C\textsuperscript{a} & 0.0796 & 173.3 & 106.6154 & 70.1956 & 51.3 & 32.6 & 44.1 & 47.4 & (34) \\
FRB20250227A\textsuperscript{a} & 0.0314 & 156.6 & 113.3792 & 15.2233 & 77.9 & 34.3 & 39.1 & 44.1 & (52) \\
FRB20250303A & 0.3058 & 489.5 & 104.7963 & 13.2836 & 149.6 & 37.3 & 50.5 & 30.0 & (34) \\
FRB20250316A\textsuperscript{a} & 0.00635 & 161.82 & 182.4347 & 58.8491 & 29.3 & 30.8 & 20.0 & 36.3 & (55) \\
FRB20250410E & 0.498 & 619.1 & 14.0417 & -5.1386 & 28.1 & 31.0 & 13.9 & 47.0 & (52) \\
FRB20250507A & 0.1204 & 212.6 & 359.6333 & 42.9422 & 69.5 & 38.0 & 44.8 & 37.0 & (52) \\
FRB20250511A & 0.218 & 360.72 & 17.3866 & -44.5272 & 28.3 & 33.2 & 9.0 & 42.0 & (50) \\
FRB20250515A\textsuperscript{a} & 0.0625 & 169.4 & 42.8708 & 8.0361 & 35.7 & 30.2 & 31.4 & 32.2 & (52) \\
FRB20250518 & 0.64 & 921.2 & 207.9918 & 71.2823 & 34.5 & 32.4 & 26.8 & 42.9 & (34) \\
FRB20250520A & 0.1685 & 341.66 & 177.7245 & -45.2137 & 110.2 & 52.5 & 47.8 & 74.2 & (50) \\
FRB20250607C & 0.2106 & 335.87 & 38.1810 & -39.3411 & 28.9 & 31.6 & 13.9 & 31.9 & (50) \\
FRB20250613A\textsuperscript{a} & 0.0987 & 174.6 & 69.1377 & -44.5329 & 38.9 & 33.1 & 28.0 & 44.7 & (53) \\
FRB20250704A & 0.113 & 356.4 & 108.4496 & 24.9602 & 81.9 & 34.0 & 38.9 & 31.5 & (52) \\
FRB20251024B & 0.319 & 543.54 & 37.3769 & -62.9628 & 35.3 & 35.6 & 18.2 & 55.0 & (50) \\
FRB20251117A & 0.2719 & 473.9 & 17.0217 & 14.8623 & 33.6 & 31.3 & 26.5 & 44.2 & (34) \\
FRB20251125A & 0.327 & 610.5 & 213.5579 & 15.3215 & 24.5 & 36.5 & 10.1 & 30.7 & (34) \\
FRB20251130A & 0.4191 & 977.0 & 158.4727 & -27.8917 & 59.7 & 39.5 & 44.7 & 58.5 & (54) \\
FRB20251225B & 0.2188 & 300.3 & 207.0951 & 17.5078 & 23.9 & 34.8 & 10.5 & 28.5 & (34) \\
FRB20260324E & 0.2475 & 343.3 & 63.8932 & 8.4432 & 48.0 & 31.3 & 42.1 & 23.1 & (50) \\
FRB20260331E & 0.21 & 307.99 & 346.7027 & -58.2013 & 35.1 & 39.7 & 13.9 & 38.1 & (50) \\
FRB20260407E & 0.4878 & 494.1 & 252.6683 & 14.5327 & 49.0 & 52.1 & 23.7 & 48.0 & (34) \\
FRB20260608D & 0.113 & 331.05 & 153.6131 & 18.3093 & 31.7 & 30.5 & 24.2 & 43.2 & (50) \\
\end{longtable}
\endgroup

\noindent\textsuperscript{a}Sources excluded from the baseline sample by the redshift cut. FRB20200120E, FRB20220319D, and FRB20230718A also fail the DM cut $\DM_{\rm exgal}>5\,\mathrm{pc\,cm^{-3}}$, evaluated with NE2025 and a Galactic halo contribution of $50\,\mathrm{pc\,cm^{-3}}$.

\noindent\textsuperscript{b}Photometric host-galaxy redshifts from Refs.~(16) and (29). We retain FRB20190614D, which is included in the sample of \citet{Sharma2026suppression}, and adopt the photometric redshift of candidate B from Ref.~(16).

\noindent\textsuperscript{c}The two candidate hosts of FRB20230311A have nearly identical spectroscopic redshifts ($z=0.1918$ and $0.1924$). We adopt $z=0.1918$ [Ref.~(37)].

\noindent\textsuperscript{d}The optical PATH posteriors for FRB20190303B and FRB20190502A are approximately $65\%$ and $45\%$, respectively. We retain their spectroscopic redshifts because both associations are supported by LoTSS radio matches, each with an estimated chance-alignment probability below $0.1\%$ [Ref.~(11)].

\noindent\textsuperscript{e}NE2025 denotes the Galactic interstellar-medium model of \citet{Ocker2026}. 

\noindent\textsuperscript{f}YT20, L26, and Y25 denote the halo prescriptions of \citet{HaloYT2020}, \citet{HaloLucchini2026}, and \citet{HaloHuang2025}, respectively. Foreground DMs are rounded to one decimal place for display. The likelihood uses their stored precision.

\noindent\textit{References.} The source numbers in the final column identify discovery, localization, and host-galaxy studies. They correspond to the following citations:

(1) \citet{FRBData103847204182138342l7}. (2) \citet{FRBData10384715384357abb6fb}. (3) \citet{FRBData10384715384357ace5aa}. (4) \citet{FRBData10384720418213aae7cb}. (5) \citet{FRBData10384715383881ac3aec}. (6) \citet{FRBData10384715384357accf89}. (7) \citet{Shannon2025}. (8) \citet{FRBData10384720418213ac223b}. (9) \citet{FRBData10384720418213ad64d1}. (10) \citet{FRBData10384715384357ad0893}. (11) \citet{Sun2026hosts}. (12) \citet{Moroianu2026}. (13) \citet{FRBData101038s41586022047555}. (14) \citet{FRBData10384715384357adb84d}. (15) \citet{FRBData101038s4158601913897}. (16) \citet{FRBData10384715384357aba4ac}. (17) \citet{FRBData10384720418213abeaa6}. (18) \citet{FRBData101038s4155002502566y}. (19) \citet{FRBData101038s4158602104354w}. (20) \citet{Shin2024sightline}. (21) \citet{FRBData101093mnrasstac1450}. (22) \citet{FRBData10384720418213ac242b}. (23) \citet{FRBData10384715384357acc178}. (24) \citet{FRBData101093mnrasstad3329}. (25) \citet{FRBData101093mnrasstad1839}. (26) \citet{FRBData101038s4155002402357x}. (27) \citet{FRBData101038s41586024080749}. (28) \citet{FRBData10384715384357ad3736}. (29) \citet{FRBData101093mnrasstaf2144}. (30) \citet{FRBData10384715383881adc725}. (31) \citet{Li2026magneto}. (32) \citet{FRBData101126scienceadf2678}. (33) \citet{FRBData101093mnrasstae1652}. (34) \citet{McCarty2026}. (35) \citet{FRBData10384720418213acc4b6}. (36) \citet{Mckinven2025}. (37) \citet{FRBData10384715384365addbda}. (38) \citet{FRBData10384715384357ae1014}. (39) \citet{FRBData101093mnrasstae2756}. (40) \citet{FRBData10384715384357ae5060}. (41) \citet{FRBData101093mnrasstaf289}. (42) \citet{FRBData10384720418213adadfd}. (43) \citet{FRBData10105100046361202453233}. (44) \citet{FRBData10384720418213ae0b68}. (45) \citet{Glowacki2025hosts}. (46) \citet{FRBData10384720418213ad9de2}. (47) \citet{FRBData10384720418213ad9ddc}. (48) \citet{Caleb2026highz}. (49) \citet{FRBData10384715384357ae0298}. (50) \citet{Dial2026HTR2}. (51) \citet{FRBData10384715384357ae2ea0}. (52) \citet{Dong2026metallicity}. (53) \citet{Dial2026repeater}. (54) \citet{Gordon2026ATel}. (55) \citet{FRB250316A2025ApJ}.

\clearpage
\renewcommand{\bibsection}{}
\bibliographystyle{apsrev4-1}
\bibliography{reference}